\documentclass[amsmath,amssymb,nofootinbib,showpacs,twocolumn,nofootinbib]{revtex4-1}
\usepackage{dcolumn}
\usepackage{bm}
\usepackage{graphicx}
\usepackage{epstopdf}
\usepackage{epsfig}
\usepackage{color}
\usepackage{wasysym}
\usepackage{natbib}
\usepackage{twoopt}
\usepackage{verbatim}
\usepackage{dashrule}
\usepackage[breaklinks=true,colorlinks=true,linkcolor=blue,citecolor=blue,urlcolor=blue,pdfauthor={Tomassetti},pdftitle={XS}]{hyperref}
\def\MyTitle#1{\section{{#1}}} 
\def\Journal#1#2#3#4{{#4}, {#1}, {#2}, #3} 

\newcommand{\GALPROP}{\textsf{GALPROP}}
\newcommand{\WNEW}{{\textsf{WNEW}}}
\newcommand{\YIELDX}{\textsf{YIELDX}}
\newcommand{\etal}{et al.}

\newcommand{\AMS}{\textsf{AMS}}
\newcommand{\XS}{\textsf{XS}}
\newcommand{\XSs}{\textsf{XS's}}
\newcommand{\eg}{\textit{e.g.}} 
\newcommand{\ie}{\textit{i.e.}} 
\newcommand{\Hyd}{\textsf{H}}

\newcommand{\He}{\textsf{He}}
\newcommand{\Li}{\textsf{Li}}
\newcommand{\Be}{\textsf{Be}}
\newcommand{\B}{\textsf{B}}
\newcommand{\C}{\textsf{C}}
\newcommand{\N}{\textsf{N}}
\newcommand{\Oxy}{\textsf{O}}

\newcommand{\BeB}{\textsf{Be}/\textsf{B}}  
\newcommand{\BC}{\textsf{B}/\textsf{C}}
\newcommand{\R}{\ensuremath{\mathcal{R}}}

\begin{document}
\title{Examination of uncertainties in nuclear data \\for cosmic ray physics with the AMS experiment}
\author{Nicola Tomassetti}
\address{LPSC, Universit\'e Grenoble-Alpes, CNRS/IN2P3, F-38026 Grenoble, France; email: nicola.tomassetti@lpsc.in2p3.fr}
\date{October 2015} 
%
%
%
\begin{abstract}
High-energy \Li-\Be-\B{} nuclei in cosmic rays are being measured with unprecedent accuracy by the \AMS{} experiment.
These data bring valuable information to the cosmic ray propagation physics. In particular, combined measurements of 
\BC{} and \BeB{} ratios may allow to break the parameter degeneracy between the cosmic-ray diffusion coefficient and 
the size of the propagation region, which is crucial for dark matter searches. 
The parameter determination relies in the calculation of the \Be{} and \B{} production from collisions of heavier 
nuclei with the gas. Using the available cross-section data, I present for the first time an evaluation of the nuclear 
uncertainties and their impact in constraining the propagation models. 
I found that the \AMS{} experiment can provide tight constraints on the transport parameters allowing to resolutely 
break the degeneracy, while nuclear uncertainties in the models are found to be a major limiting factor.
Once these uncertainties are accounted, the degeneracy remains poorly resolved. 
In particular, the \BeB{} ratio at $\sim$1\,-\,10 GeV/n is found not to bring valuable information for 
the parameter extraction. On the other hand, precise \BeB{} data at higher energy may be useful
to test the nuclear physics inputs of the models. 
\end{abstract}
\maketitle

\MyTitle{Introduction}  
%
Understanding the cosmic ray (CR) transport processes in the Galaxy is a major subject in modern astrophysics.
The CR transport is studied using data on secondary nuclei (\eg, \Li-\Be-\B), which are created by 
fragmentation of heavier elements, and primary nuclei (\eg, \Hyd, \He, \C-\N-\Oxy), which are produced and 
accelerated in Galactic sources.
Secondary-to-primary ratios, and in particular the \BC{} ratio,
are used to constrain the Galactic diffusion coefficient, $D$, and the half 
vertical extent of the propagation region, $L$ \citep{Grenier2015,Strong2007,Maurin2001}.
For a rigidity-dependent diffusion coefficient $D \approx D_{0}\R^{\delta}$, the \BC{} ratio fixes both $\delta$ and the $D_{0}/L$ ratio.
The degeneracy between $D_{0}$ and $L$ may be lifted using data with unstable isotopes, such as the $^{10}$\Be/$^{9}$\Be{} 
isotopic ratio or the \BeB{} elemental ratio \citep{WebberSoutoul1998,Putze2010}. 
Hence, combined measurements of \BC{} and \BeB{} ratios may provide the determination of the basic CR transport parameters.
Understanding CR transport is crucial to reliably 
predict the secondary production of antimatter and to set stringent
limits on dark matter annihilation signals. 
The parameter $L$ is of great importance for assessing the dark matter signal.

The spectra of \B{} and \Be{} nuclei are now being measured by the Alpha Magnetic Spectrometer 
(\AMS) experiment in the International Space Station (ISS).
Recent measurements of CR protons \citep{Aguilar2015} and preliminary results on light nuclei \citep{Oliva2015,Derome2015}
show that \AMS{} is probing the GeV-TeV energy region to a  $\sim$\,\% level of accuracy.
With this standard of precision, it is now timely to 
assess the theoretical uncertainties of the model predictions.
In particular, calculations of \Be-\B{} production rates rely on several cross-section (\XS) estimates. 
Propagation models make use of semi-empirical \XS{} formulae calibrated to accelerator data. 
Thus, the accuracy of the inferred transport parameters is directly linked to the quality of the available measurements on nuclear fragmentation.

The aim of this paper is 
to estimate the \emph{nuclear uncertainties} in CR propagation and to investigate how they affect the 
determination of the CR transport parameters. In particular, I will focus on the anticipated \AMS{} data 
on the \BC{} and \BeB{} ratios and their connection with the $D_{0}/L$ degeneracy problem.
For this purpose, I have gathered all available \XS{} data for \Be{} and \B{} production from \B-\C-\N-\Oxy{} 
collisions off hydrogen target. These data have been used to constrain the \XS{} parameterizations in order to 
estimate their uncertainties. The resulting \XS{} errors have been therefore converted into theoretical uncertainties 
for the model predictions and, finally, into uncertainties on the transport parameters that can potentially be inferred by \AMS{}.

\MyTitle{Calculations} 
%
\emph{The CR propagation model} --- 
This work relies on the diffusive-reacceleration model implemented under
the code \GALPROP, 
which numerically computes the equilibrium spectra of CR leptons and nuclei 
for given source functions and boundary conditions \citep{StrongMoskalenko1998,Trotta2011}.
I define a \emph{reference model} as follows. The source spectra are taken as broken power-law
functions, $q_{j} \propto (\R/\R_{\B})^{-\nu}$, with index $\nu_{1}=1.9$ ($\nu_{2}=2.38$) below (above) $\R_{\rm B}=$\,9\,GV.
The diffusion coefficient is taken as $D(\R)=\beta D_{0}\left(\R/\R_{0}\right)^{\delta}$, 
with $D_{0}=5 \cdot\,10^{28}$\,cm$^{2}$\,s$^{-1}$, $\delta=$\,0.38, and $\R_{0}=$\,4\,GV.
The Alfv\'en speed is $v_{A}=$\,33\,km\,s$^{-1}$. 
The cylindrical diffusion region has radius $r_{max}=$\,30\,kpc and half-height $L=$\,3.9\,kpc. 
A large nuclear reaction chain is set up, describing the production 
of secondary $j$-type nuclei from fragmentation of heavier
$k$-type nuclei. The fragmentation rate is  $\Gamma_{k\rightarrow j}=\beta_{k} c n \sigma^{i}_{k\rightarrow j}(E)$,
where $n_{i}$ are the number densities of the ISM nuclei ($n_{\rm H}\cong$\,0.9\,cm$^{-3}$ 
and $n_{\rm He}\cong$\,0.1\,cm$^{-3}$)  
and $\sigma_{k \rightarrow j}^{i}$ is the production \XS{} off $i$-type target at energy $E$. 
Under \GALPROP, production \XSs{} are evaluated from interpolation/fits to the data or from 
nuclear codes, such as \texttt{CEM2k} or \texttt{LAQGSM}, eventually normalized to the 
data \citep{Gudima1983,Gudima2001,Mashnik1998,MoskalenkoMashnik2003}. 
%
\begin{figure} 
\includegraphics[width=0.46\textwidth]{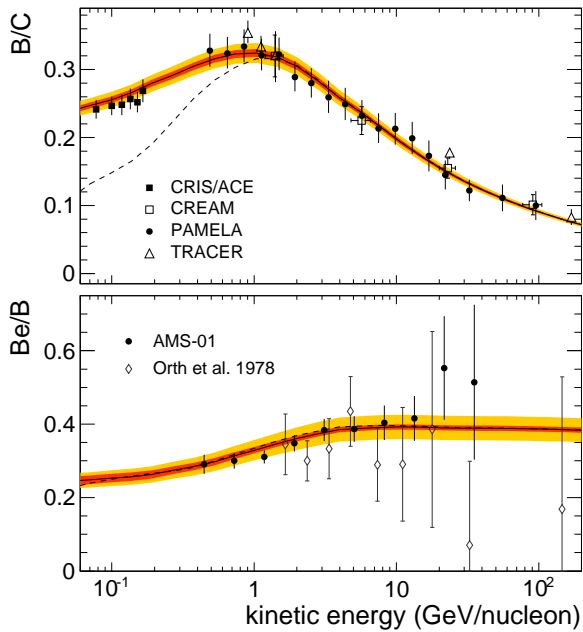}
\caption{\footnotesize%
  (Color online) Elemental ratios \BC{} (top) and \BeB{} (bottom) from the reference model in comparison
  with the data \cite{AMS01Nuclei2010,Adriani2014,Ahn2008,Obermeier2012,DeNolfo2003,Orth1978}.  
  The yellow bands are the estimated nuclear uncertainties (from Fig.\,\ref{Fig::ccCSBoronProd}).
  The red bands reflect the estimated parameter uncertainties 
  for anticipated \AMS{} data (from Fig.\,\ref{Fig::ccChiSquaresAMS02}).
}\label{Fig::ccBenchmarkModel}
\end{figure}
%
\begin{figure*} 
  \includegraphics[width=0.94\textwidth]{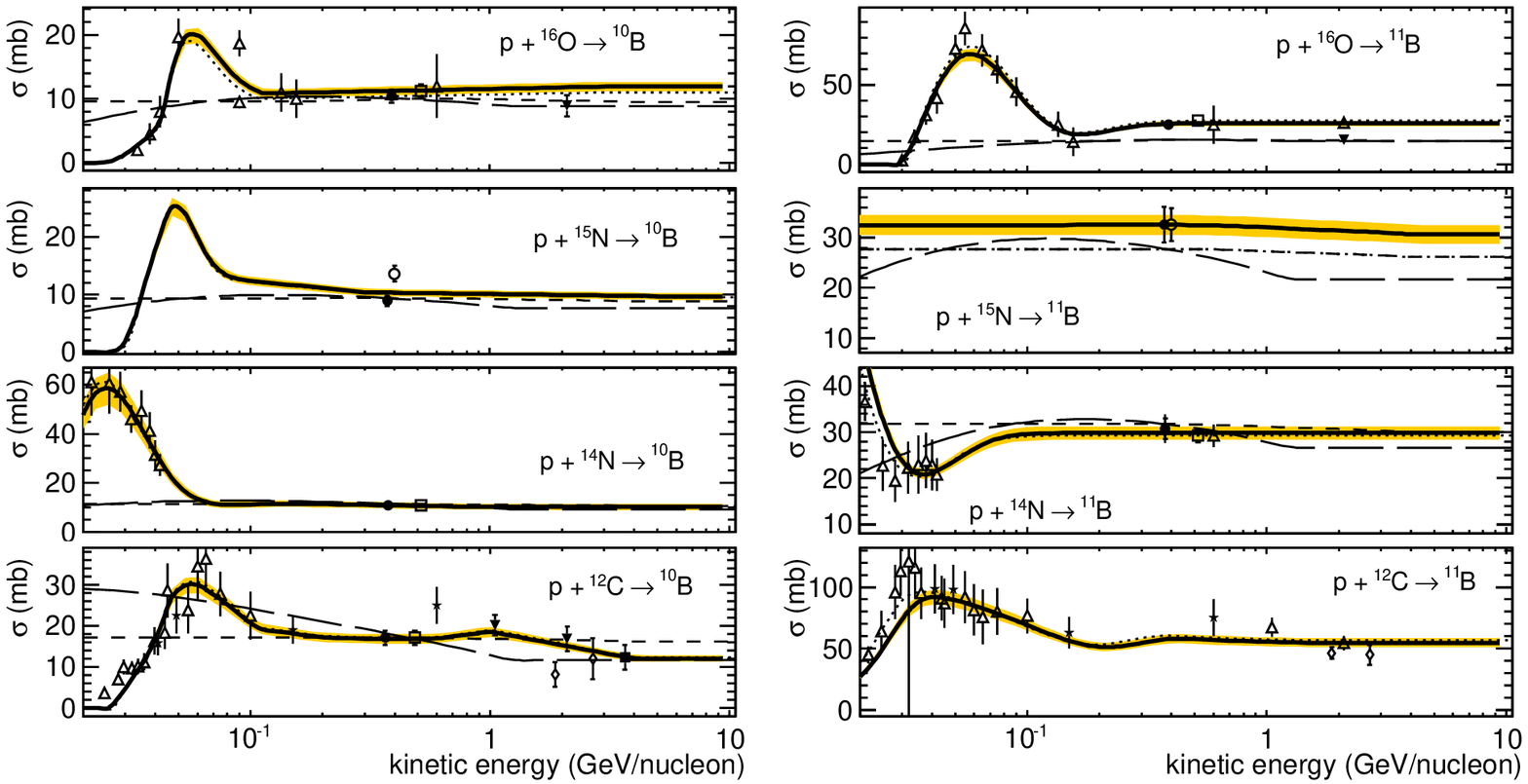} \label{Fig::ccCSBoronProd}
\qquad
  \includegraphics[width=0.94\textwidth]{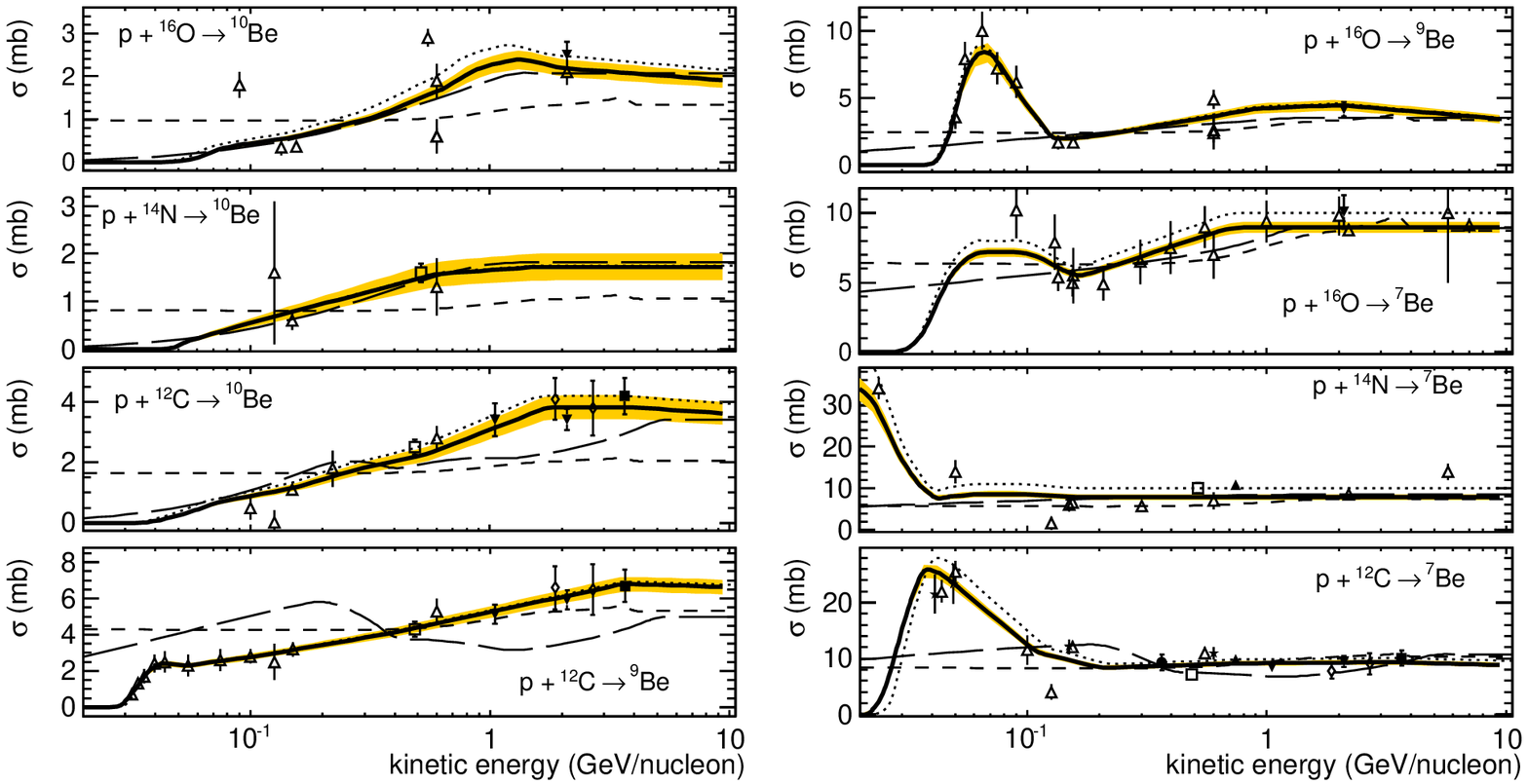}
\qquad
  \includegraphics[width=0.94\textwidth]{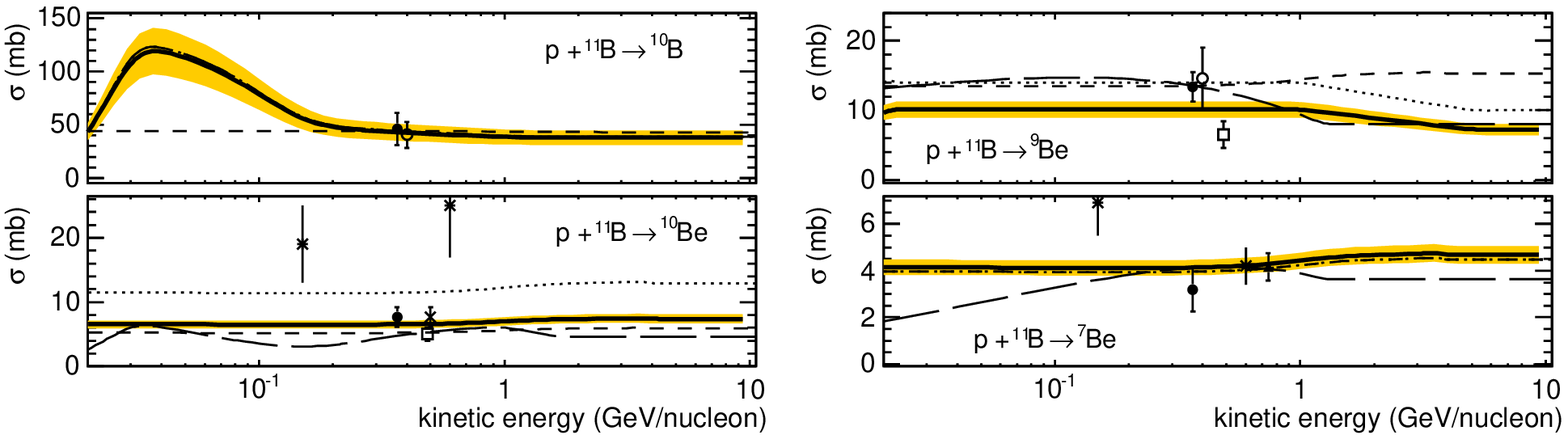}
  \caption{\footnotesize%
    (Color online) Fragmentation \XSs{} for $^{10}$\B, $^{11}$\B, $^{7}$\Be{}, $^{9}$\Be, and $^{10}$\Be{} production from 
    \B-\C-\N-\Oxy{} collisions off hydrogen. The data are from
    \citet{ReadViola1984,Webber1998,Webber1990c,Olson1983,Fontes1977,Korejwo2000,Korejwo2001,Radin1979,Ramaty1997,Webber1998prc,Raisbeck1971}.
    The lines are from the \WNEW{} (short-dashed), \YIELDX{} (long-dashed), \GALPROP{} (dotted), 
    and the \XSs{} determined in this work (thick solid lines) with their uncertainty band.
  }\label{Fig::ccCSBoronProd}
\end{figure*}

\emph{Nuclear uncertainties} --- 
%
Data on isotopically separated targets and fragments have been collected by several experiments, 
but only in narrow energy ranges. A compilation is shown in Fig.\,\ref{Fig::ccCSBoronProd}
for the production of $^{10}$\B, $^{11}$\B, $^{7}$\Be, $^{9}$\Be, and $^{10}$\Be{} isotopes
from fragmentation of $^{12}$\C, $^{14,15}$\N{} and $^{16}$\Oxy{} off hydrogen target at 
energy between 30\,MeV/n and 10\,GeV/n. 
For \Be{} production, I have also considered \textit{tertiary} reactions such as \B$\rightarrow$\Be{}, \ie,
those reactions where the progenitor nuclei are of secondary origin.
All these processes account for $\gtrsim$\,90\% of the \Be{} and \B{} production.
At energy above than a few GeV/n, all the \XSs{} are nearly constant in energy. 
The data in Fig.\,\ref{Fig::ccCSBoronProd} are compared with \XS{} parameterizations from \GALPROP{}
and from the popular formulae \WNEW{} \citep{Webber1990b,Webber1990d,Webber2003} 
and \YIELDX{} \citep{Silberberg1998}.
Following earlier studies \citep{Tomassetti2012,Moskalenko2011}, the \XS{} uncertainties 
are determiend by a re-fit of the \GALPROP{} parameterizations $\sigma_{\rm G}(E)$ to the data.
For each reaction, the corresponding \XS{} are fit with the function 
$\sigma_{\rm H}(E) = a \sigma_{\rm G}(b E)$, 
where $a$ and $b$ are free parameters representing normalization and energy scale. 
This procedure allows to determine a new set of \XSs{} and their associated uncertainties 
corresponding to one-sigma confidence intervals. The uncertainty bands are shown in Fig.~\ref{Fig::ccCSBoronProd}. 
The new \XSs{} are often close to the original $\sigma_{\rm G}$ values,
but the \Be{} production in \GALPROP{} is found to be over-estimated by a few percent. 
Such a \Be{} overproduction was also reported in \citet{AMS01Nuclei2010}.
\begin{figure*}[!t] 
\includegraphics[width=0.89\textwidth]{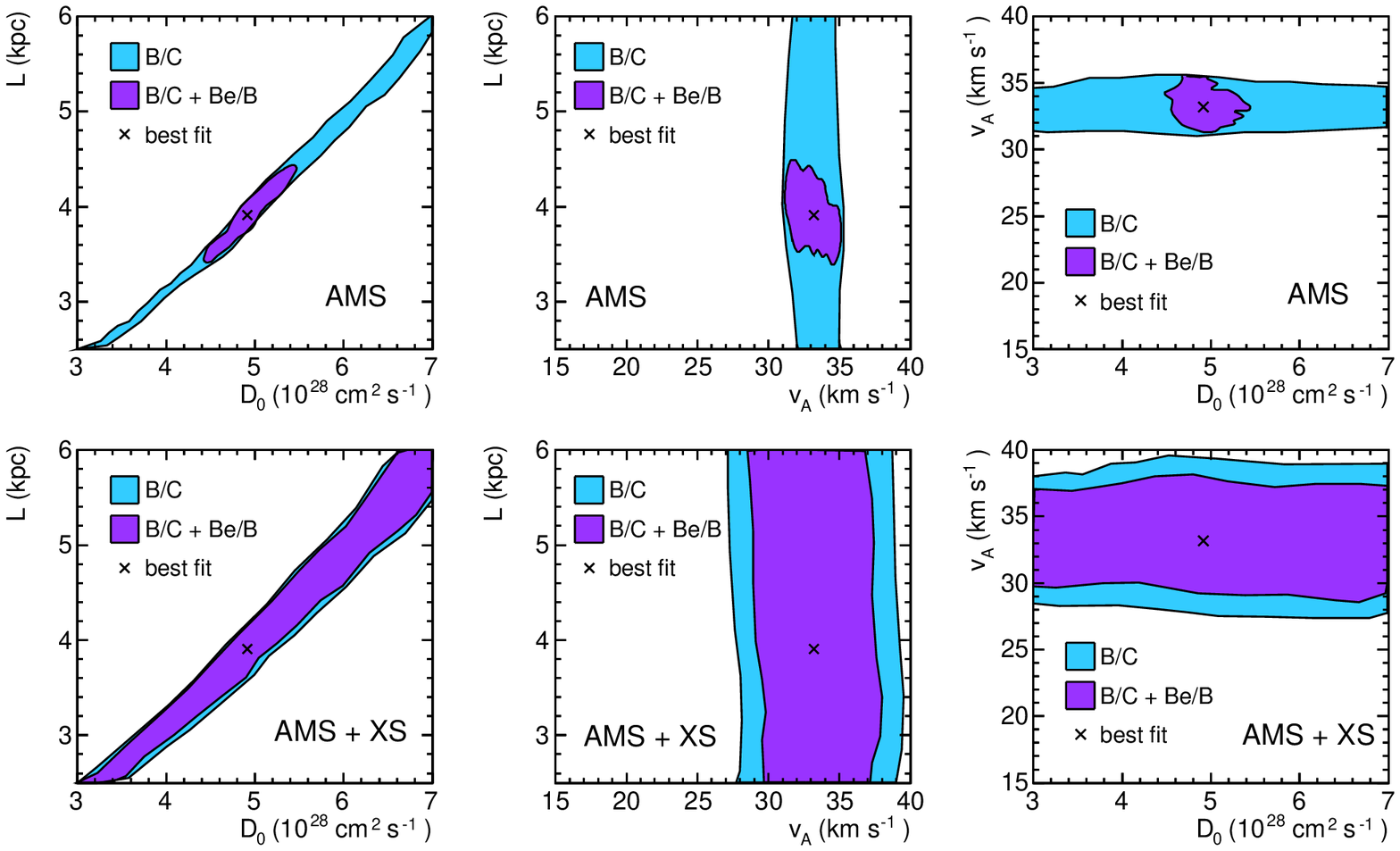}
\caption{\footnotesize%
  (Color online) Top: estimation of the \AMS{} capabilities in constraining the parameters 
  $D_{0}$, $L$, and $v_{A}$ with with the \BC{} and \BeB{} ratios.
  Bottom: same as above after accounting for \emph{nuclear uncertainties} in the \Be-\B{} production rates.
}\label{Fig::ccChiSquaresAMS02}
\end{figure*}
%
The estimated \XS{} errors have been converted into uncertainties of the secondary/tertiary production terms and 
then propagated at Earth. Typical uncertainties are found to be $\sim$\,5\,\% for \B{} production and 
$\sim$\,7\--10\% for \Be{} production, with $\sim$\,10\% for $^{10}$\Be{} productions. 
The corresponding uncertainties in the \BC{} and \BeB{} ratios are shown in Fig.\,\ref{Fig::ccBenchmarkModel} 
for the reference model (yellow bands). 

\emph{Modeling the \AMS{} performance} --- 
%
I consider the \BC{} ratio at 2 - 200\,GeV/n and the \BeB{} ratio at 1 - 100\,GeV/n. 
For these ratios, I compute the anticipated \AMS{} data under the reference model.
The number of $j$--type particles detected by \AMS{} in each energy bin is estimated
via the convolution of the CR flux with the detector acceptance \citep{TomassettiDonato2012},
$\Delta N_{j} = \int \Phi_{j} \mathcal{G}_{j} \mathcal{T}_{j} dE$,
where $\Phi_{j}$ is the input spectrum, $\mathcal{G}_{j}$ is the total detector acceptance
and $\mathcal{T}_{j}$ is the effective exposure time for a total data taking period $T_{0}$.
All input spectra are solar-modulated under the force-field approximation \citep{Gleeson1968},
using $\phi\cong 550$\,MV for the \AMS{} observation period. 
I consider the case of 10 bins per decade, log-uniformly spaced in energy, 
and a total exposure of $\mathcal{G} {T_{0}} \cong$\,100\,m$^{2}$\,sr\,day. 
The effective exposure time must account for the geomagnetic field modulation which 
suppresses the Galactic CR flux below
the cut-off rigidity, $\R_{\rm C}\approx$\,0.5--20\,GV, depending on the detector location.
I adopt the St\"ormer model, $\R_{C}(t)={20\,{\rm GV}}{\rho^{-1}(t)}cos^{4}\theta_{\rm M}(t)$ 
\citep{SmartShea2005}, where $\theta_{\rm M}(t)$ is the geomagnetic latitude and $\rho(t)$ is 
the distance between \AMS{} and the geomagnetic dipole center in units of Earth's radii. 
Their evolution depends on the ISS orbit around the Earth.
The function  $\mathcal{T}_{j}$ is computed as $\mathcal{T}_{j}(E)= \int_{T_{0}} \alpha(t) \mathcal{H}_{j}(t,E) dt$,
where $\alpha\approx$\,95\,\% is the detector live-time,
and $\mathcal{H}_{j}(t,E)$ is the geomagnetic transmission function, which is modeled as
an $\R$-dependent smoothed step function, $\mathcal{H}=\left[ 1 + (\R/\R_{C})^{-12} \right]^{-1}$.
Its particle-dependence arises from the conversion $\R\rightarrow E$, while
its time-dependence is contained in $\R_{C}(t)$.
The integral $\mathcal{T}_{j}(E)$ has been numerically computed for all relevant isotopes 
by simulating 23,000 ISS orbits with period $T_{\rm ISS}=91\,$\,min and inclination $\theta_{\rm ISS}=\,$51.7$^{\circ}$. %
Systematic errors are assigned to be 1.5\,\% for the \BC{} ratio and 1\,\% on the \BeB{} ratio,
constant in the considered energy range \citep{Oliva2015}.
From the estimated counts, the statistical errors associated with the \BC{} ratio are given 
by $1/\sqrt{\Delta N_{\rm B}} + 1/\sqrt{\Delta N_{\rm C}}$, and similarly for the \BeB{} ratio.
%
%
\MyTitle{Results and discussions} 
%
\emph{The \AMS{} physics potential} --- 
The \AMS{} capabilities in constraining the model parameters are first estimated \emph{without} accounting for nuclear uncertainties.
For this purpose, I have performed a scan in the parameter space $D_{0} \times L \times v_{A}$ 
by running \GALPROP{}  3,420 times over a $19 \times 15 \times 12$ grid. 
The resulting spectra are then tri-linearly interpolated to a $4\times$ finer parameter grid, corresponding to 187,245 models.
Hence, the \BC{} ratio predicted by each model is tested against the artificial \AMS{} data 
(generated with the reference model of Fig.~\ref{Fig::ccBenchmarkModel}), using the $\chi^{2}$ method. 
The same procedure is done for \BeB{} ratio and for the two combined ratios.
The results of are shown in Fig.~\ref{Fig::ccChiSquaresAMS02}, top panels, 
where the one-sigma contour regions are plotted as 2D projections of the parameter space.
These regions are obtained from the $\chi^{2}$ surfaces of the \BC{} ratio and 
of the \BC{}$+$\BeB{} combination.
The best-fit model on each plot (marked as ``$\times$'') always recovers the true reference model. 
The complementarity of the two ratios in breaking the $L$-$D_{0}$ degeneracy is apparent.
While the \BC{} ratio constrains the two parameters into a tight region of the ($L,D_{0}$) plane, 
only the combination \BC{}+\BeB{} allows to resolve their single values.
The Alfv\'en speed $v_{A}$ is well determined by means of \BC{} data only.
Tighter constraints be obtained using data below 2\,GeV/n, 
provided that the solar modulation effect is well modeled.
The accuracy of the measured parameters is $\delta D_{0}\sim\,0.5 \cdot 10^{28}$\,cm$^{2}$s$^{-1}$, $\delta L\sim\,0.5$\,kpc, 
and $\delta v_{A}\,\sim 2$\,km/s.
This level of accuracy, from the estimated \AMS{} capability, would represent quite a significant progress in CR propagation. 

\emph{Impact of nuclear uncertainties} --- 
The models constrained by \AMS{} are shown in Fig.\,\ref{Fig::ccCSBoronProd}
for both ratios (red bands).
As clear from the figure, nuclear uncertainties (yellow bands) are dominating.
In order to evaluate how these uncertainties affect the parameter reconstruction,
I have repeated the parameter determination procedure after accounting for the \XS{} errors in 
the $\chi^{2}$ calculations. 
The results are shown in the bottom panels of Fig.\,\ref{Fig::ccChiSquaresAMS02}. 
In comparison with the top panels, one can see that nuclear uncertainties have a 
dramatic impact on the parameters $D_{0}$ and $L$.
As shown in the figure, the $D_{0}/L$ degeneracy remain essentially unresolved 
when the nuclear uncertainties are taken into account. 
In fact, the information needed to break the $D_{0}/L$ degeneracy is contained in 
the $^{10}$\Be$\rightarrow$$^{10}$\B{} decay
which produces only small variations in the \BeB{} ratio.
Along with large uncertainties on the $^{10}$\Be{} production, this information is also 
washed out by uncertainties in the more abundant $^{7,9}$\Be{} and $^{11}$\B{} components of the \BeB{} ratio. 
At this point one may argue that a direct, ideal measurement of $^{10}$\Be{} at $\sim$\,1--10\,GeV/n would bring tighter constraints. 
Thus, I have repeated the calculations after considering \XS{} uncertainties 
for the $^{10}$\Be{} production only, \ie, 
assuming ideal knowledge of the other isotopes and infinite precision measurements. 
The sole uncertainties in the $^{10}$\Be{} production would
limit the parameter reconstruction to $\delta D_{0} \sim 1.5 \cdot 10^{28}\,$cm$^{2}$s$^{-1}$ and $\delta L \sim 1.5\;$kpc. 
This still represents a poor parameter determination in comparisons to the \AMS{} potential.
Nonetheless, given the current level of nuclear uncertainties, 
a direct measurement of $^{10}$\Be{} flux (even if affected a few $\%$ systematic errors)
would probably bring better information than a precise \BeB{} measurement.
\begin{figure}[!t] 
\includegraphics[width=0.46\textwidth]{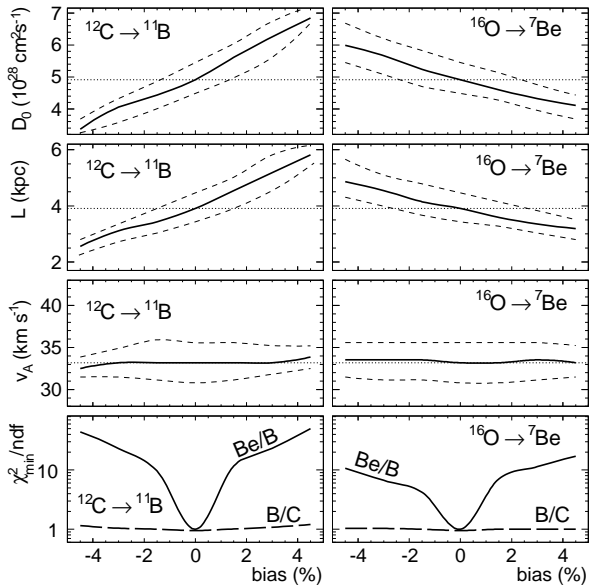}
\caption{\footnotesize%
  (Color online) Best-fit parameters $D_{0}$, $L$, and $V_{A}$ and the corresponding $\chi^{2}_{\rm min}$ as a function of the 
  relative bias introduced in the \XS{} normalization for the reactions  $^{12}$\C$\rightarrow$$^{11}$\B{} and $^{16}$\Oxy$\rightarrow$$^{7}$\Be. 
}\label{Fig::ccBias}
\end{figure}

\emph{Single-reaction \XS{} bias} --- 
%
It is instructive to study the dependence of the best-fit parameters on single \XS{} reactions.
An example is the anti-correlation between the reaction 
$^{11}$\B$\rightarrow$$^{10}$\Be{} and the parameter $L$ \cite{Hams2004}. 
Here I consider the reactions $^{12}$\C$\rightarrow$$^{11}$\B{} and $^{16}$\Oxy$\rightarrow$$^{7}$\Be. 
After the introduction of systematic biases in their \XS{} normalizations, 
I have generated new reference-model predictions
and new \AMS{} mock data. Then, I have repeated the parameters 
reconstruction procedure using the nominal set of unbiased \XSs.
The results are shown is in Fig.~\ref{Fig::ccBias}. The best-fit parameters
are plotted as a function of the bias induced on the \XS{} normalization for the considered reactions. 
The horizontal dotted lines show the true parameter values.
In this case, the best-fit parameters are not correctly reconstructed. 
While the determination of ${v}_{A}$ appears rather stable, 
the parameter ${L}$ and ${D}_{0}$ are mis-inferred by large factors
after the introduction of a few \% deviation in the relevant {\XS}s. 
This shows, again, that the constraining power of the \BeB{} ratio is weakened by uncertainties in the 
$^{11}$\B{} or $^{7,9}$\Be{} production, rather than by the large errors on the $^{10}$\Be{} production.
It is also interesting to look at the evolution of the best-fit $\chi_{\rm min}^{2}$ 
as function of the bias. This is shown in the bottom panels of the figure.
The best $\chi^{2}$ of the \BC{} ratio appears insensitive to \XS{} biases.
In fact the \BC{} ratio is approximately given by $\BC \propto \Gamma_{B}/(D/L)$, 
\ie, any deviation in the \B{} production rate $\Gamma_{B}$
can be re-absorbed by a different determination of $D_{0}/L$.
This is not the case for the \BeB{} ratio
because its high-energy plateau (at $\gtrsim$\,10\,GeV/n) 
is almost independent on propagation effects. 
Hence, discrepancies between \BeB{} data and model predictions cannot 
be re-absorbed by the fit in term of different parameter combinations: 
they can only arise from nuclear physics inputs. 
An example of this is found in \citet{AMS01Nuclei2010}, where the small discrepancy between the \BeB{} data and the model 
predictions was ascribed to the \XSs{} for \Be{} production.
The \BeB{} ratio, like other secondary-to-secondary ratios \citep{AMS01Isotopes2011}, may be therefore used
as a diagnostic tool to detect possible biases in the production \XSs.

\MyTitle{Conclusions} 
%
My estimates show that the \AMS{} experiment can provide tight constraints on the key 
parameters $D_{0}$, $L$, and $v_{A}$.
Given the level of precision expected by \AMS, nuclear uncertainties in secondary production 
models are found to be a major limitation in the interpretation of secondary CR nuclei. 
Once nuclear uncertainties are accounted, 
the $D_{0}/L$ degeneracy remains poorly resolved. 
With the current status of nuclear data, the \BeB{} ratio appears not to bring 
valuable information for the parameter extraction. 
Isotopically resolved CR measurements such as the $^{10}$\Be/$^{9}$\Be{} ratio are preferable, 
though the $^{10}$\Be{} production rate is affected by large uncertainties.
On the other hand, precise data on the \BeB{} ratio at $E\gtrsim$\,10\,GeV/n
may represent a powerful tool to test the nuclear physics inputs of the propagation models, 
and in particular to detect possible biases 
in single reactions that may cause a parameter mis-determination.
It also worth stressing that this problem has a direct impact in dark matter searches. 
My study provides a concrete case study for Gondolo's plea to the nuclear physics community \citep{Gondolo2014}. 
In summary, nuclear uncertainties are a major limiting factor for further progress in CR propagation.
The collection of new nuclear data, within a dedicated program of \XS{} measurements and modeling, 
would enable to fully exploit the potential of the \AMS{} data.
\\

\footnotesize{
I thank I. Gebauer, F. Donato, L. Derome, D. Maurin, A. Oliva
for discussion, I. Moskalenko and the \GALPROP{} team for sharing 
their code with the community. 
This work is supported by the ANR LabEx grant  \textsf{ENIGMASS} at CNRS/IN2P3.
}

\end{document}